# Reflection of an ultrasonic wave on the bone-implant interface: effect of the roughness parameters

Yoann Hériveaux[1], Vu-Hieu Nguyen[2], Vladimir Brailovski[3], Cyril Gorny[4], Guillaume Haïat[1]

[1]: CNRS, Laboratoire Modélisation et Simulation Multi-Échelle, MSME UMR 8208 CNRS, 61 avenue du Général de Gaulle, 94010 Créteil Cedex, France.
[2]: Université Paris-Est, Laboratoire Modélisation et Simulation Multi Echelle, MSME UMR 8208 CNRS, 61 avenue du Général de Gaulle, 94010 Créteil Cedex, France.
[3]: Department of Mechanical Engineering, École de technologie supérieure, 1100 Notre-Dame Street West, Montreal, QC H3C 1K3, Canada;
[4]: Laboratoire PIMM (ENSAM, CNRS, CNAM, Hesam Université), 151 Bd de l'Hôpital 75013 Paris, France (ENSAM)

Corresponding author: Guillaume HAÏAT
Laboratoire Modélisation Simulation Multi-Échelle, UMR CNRS 8208,
61 avenue du général de Gaulle, 94010 Créteil, France
e-mail : guillaume.haiat@univ-paris-est.fr

**Abstract –**

Quantitative ultrasound can be used to characterize the evolution of the bone-implant interface (BII), which is a complex system due to the implant surface roughness and to partial contact between bone and the implant. The aim of this study is to derive the main determinants of the ultrasonic response of the BII during osseointegration phenomena. The influence of i) the surface roughness parameters and ii) the thickness $W$ of a soft tissue layer on the reflection coefficient $r$ of the BII was investigated using a two-dimensional finite element model.

When $W$ increases from 0 to 150 μm, $r$ increases from values in the range [0.45; 0.55] to values in the range [0.75; 0.88] according to the roughness parameters. An optimization method was developed to determine the sinusoidal roughness profile leading to the most similar ultrasonic response for all values of $W$ compared to the original profile. The results show that the difference between the ultrasonic responses of the optimal sinusoidal profile and of the original profile was lower to typical experimental errors. This approach provides a better understanding of the ultrasonic response of the BII, which may be used in future numerical simulation realized at the scale of an implant.

## I. Introduction

The clinical success of endosseous implant surgery is strongly dependent on osseointegration phenomena (Khan *et al.*, 2012). The biological tissues surrounding an implant are initially non-mineralized and may thus be described as a soft tissue (Moerman *et al.*, 2016). During normal osseointegration processes, periprosthetic bone tissue is progressively transformed into mineralized bone, which may then be described as a solid. However, in cases associated to implant failures, the aforementioned osseointegration phenomena do not occur in an appropriate manner, leading to the presence of fibrous tissue around the implant. Osseointegration failure then leads to the implant aseptic loosening, which is one of the major causes of surgical failure and which remains difficult to anticipate (Pilliar *et al.*, 1986). The evolution of the implant biomechanical stability is the main determinant of the surgical success (Mathieu *et al.*, 2014) and is directly related to the biomechanical properties of the bone-implant interface (BII) (Franchi *et al.*, 2007; Mathieu *et al.*, 2014).

Various techniques such as impact methods (Schulte *et al.*, 1983; Van Scotter and Wilson, 1991 ; Mathieu *et al.*, 2013; Michel *et al.*, 2016) or resonance frequency analysis (Meredith *et al.*, 1996; Georgiou and Cunningham, 2001; Pastrav *et al.*, 2009) have been applied to investigate the BII properties. In an *ex vivo* study using a coin-shaped implant model (Vayron *et al.* (2012a), the reflection coefficient of a 15 MHz ultrasonic wave interacting with the BII significantly decreases as a function of healing time, which may be explained by a decrease of the gap of acoustical properties at the BII related to a combined increase of i) the bone-implant contact (BIC) ratio, ii) the bone Young's modulus (Vayron *et al.*, 2012a) and iii) bone mass density (Mathieu *et al.*, 2011b; Vayron *et al.*, 2014b). These results open new paths in the development of quantitative ultrasound (QUS) methods that had been previously suggested to assess dental implant stability (de Almeida *et al.* (2007). Recently, a 10 MHz QUS device was validated first *ex vivo* using cylindrical implants (Mathieu *et al.*, 2011b), then *in vitro* using dental implant inserted in a biomaterial (Vayron *et al.*, 2013) and bone tissue (Vayron *et al.*, 2014a) and, eventually, *in vivo* (Vayron *et al.*, 2014c). The sensitivity of this QUS device on changes of the periprosthetic bone tissue was shown to be significantly higher compared to the resonance frequency analysis *in vitro* (Vayron *et al.*, 2018b) and *in vivo* (Vayron et al., 2018a). These last results may be explained by a better resolution of the QUS device to changes of periprosthetic bone tissue compared to vibrational approaches.

Ultrasound techniques are also employed to stimulate bone remodeling and osseointegration through low intensity pulsed ultrasound (LIPUS) (Duarte, 1983; Tanzer *et al.*, 2001; Nakanishi *et al.*, 2011). However, the precise mechanism of action of LIPUS remains poorly understood (Claes and Willie, 2007; Padilla *et al.*, 2014), in particular because the phenomena determining the interaction between an ultrasonic wave and the BII still remain unclear. A better understanding of the aforementioned phenomena could thus help improving the performances of both QUS and LIPUS techniques. However, the various parameters influencing the interaction between an ultrasonic wave and the BII (such as periprosthetic bone quality and quantity) are difficult to control when following an experimental approach and may vary in parallel. Therefore, acoustical modeling and numerical simulation are useful because the influence of the implant and bone mechanical and geometrical properties can be precisely assessed.

A 2-D finite difference time domain (FDTD) method (Mathieu *et al.*, 2011a) and 3-D axisymmetric finite element model (FEM) have been used to model the ultrasonic propagation in a cylindrical implant (Vayron *et al.*, 2015) and in a model considering a more realistic geometry of a dental implant (Vayron *et al.*, 2016). However, the aforementioned studies considered a fully-bonded BII and did not account for the combined effect of the surface roughness and bone ingrowth around the implant. Since osseointegration was only modeled through variations of the biomechanical properties of periprosthetic bone tissue, the influence of the BIC ratio could not be considered either. More recently, a 2-D FEM has been developed to investigate the sensitivity of the ultrasonic response to multiscale surface roughness properties of the BII and to osseointegration processes (Heriveaux *et al.*, 2018). The implant roughness was modeled by a simple sinusoidal profile and the thickness of a soft tissue layer comprised between the bone and the implant was progressively reduced to simulate osseointegration phenomena. Although the sinusoidal description of the surface profile may be adapted at the macroscopic scale because it is close to mimicking implant threading, it constitutes a strong approximation in the microscopic case because the surface roughness has random characteristics.

The aim of this paper is to model the interaction between an ultrasonic wave and a rough BII considering actual surface roughness. Another related aim is to determine to what extent actual implant roughness could be replaced by a sinusoidal profile. To do so, a 2-D time domain finite element model was used to model the interaction between an ultrasonic wave and the BII.

## II. Material and methods

### A. Description of the numerical model

The numerical model considered herein was similar to the one employed in (Heriveaux *et al.*, 2018), except that actual implant surface profiles are considered. Briefly, two coupled 2-dimensional half-spaces were separated from each other by an interphase. The first domain corresponds to the implant made of titanium alloy (Ti-6Al-4V, noted (1) in Fig. 1) and the other one represents bone tissue (noted (3) in Fig. 1). Two different geometrical descriptions of the implant surface profile were considered in this study. First, the implant surface profile was defined by the results obtained using profilometry measurements (see subsection II.D), as shown in Fig. 1a. Such roughness profiles are described as "original" in what follows. Second, similarly to what has been done in (Heriveaux *et al.*, 2018), the implant surface profile was defined by a sinusoidal function of amplitude $h$ and half-period $L$, as shown in Fig. 1b. Only a single half-sine period of the interface was considered, which is sufficient to simulate the propagation of the acoustic wave using symmetrical boundary conditions for the interfaces perpendicular to the direction $x$. Note that in the case of sinusoidal surface profiles, we have: $h = \pi R_a$ and $L = S_m/2$, where $R_a$ is the arithmetical mean roughness value and $S_m$ is the mean value of the spacing between profile irregularities. The average altitude of the surface roughness was taken as the origin for the $y$ coordinates in all cases.

A soft tissue layer was considered between bone and the implant (noted (2) in Fig. 1) in order to model non-mineralized fibrous tissue that may be present at the BII just after surgery or in the case of non-osseointegrated implants (Heller and Heller, 1996). The thickness $W$ of the soft tissue layer was defined as the distance between the highest point of the surface profile and the bone level, as shown in Fig. 1. A progression of osseointegration is associated to a decrease of the value of $W$.

The total lengths of the implant and the bone domain in the direction of propagation $y$, denoted respectively $H_{Ti}$ and $H_b$, were chosen equal to 1.5 cm in order to clearly distinguish the echo reflected from the interphase and to avoid any reflection from the boundary of the simulation domain.

All media were assumed to have homogeneous isotropic mechanical properties. The values used for the different media are shown in Table I and were taken from (Njeh *et al.*, 1999; Pattijn *et al.*, 2006; Padilla *et al.*, 2007; Pattijn *et al.*, 2007). A shear wave velocity equal to 10 m/s was considered for non mineralized bone tissue following the values taken from the literature (Madsen *et al.*, 1983; Sarvazyan *et al.*, 2013) for soft tissues. However, to the best of our knowledge, the value of the shear wave velocity of non mineralized bone tissue is unknown.

The acoustical source was modeled as a broadband ultrasonic pulse with a uniform pressure $p(t)$ applied at the top surface of the implant domain (see Fig. 1) defined by:

$$p(t) = A\, e^{-4\,(f_c t - 1)^2} \sin(2\pi\, f_c\, t), \qquad (1)$$

where $A$ is an arbitrary constant and $f_c$ is its central frequency, which was set to 10 MHz throughout the study as it corresponds to the value used in the QUS device developed by our group (Mathieu *et al.*, 2011b; Vayron *et al.*, 2013; Vayron *et al.*, 2014a; Vayron *et al.*, 2014c; Vayron *et al.*, 2018a; Vayron *et al.*, 2018b). Moreover, the results obtained in (Heriveaux *et al.*, 2018) indicate that using a frequency equal to 10 MHz guarantees an acceptable sensitivity of the ultrasound response on changes of the biomechanical properties of the BII (a resolution of around 2-12 μm depending on the implant roughness was obtained).

The governing equations have been described in details in (Heriveaux *et al.*, 2018) and the reader is referred to this publication for further details. Briefly, the classical equations of elastodynamic wave propagation in isotropic solids were considered. The continuity of the displacement and stress fields were considered at each interface ($i$ - $j$), where $\{i,j\} = \{1,2\}$ or $\{2,3\}$. At the top boundary of the implant domain (at $y = H_{Ti}$, see Fig. 1b), a uniform pressure $p(t)$ was imposed. At the bottom boundary of the bone domain (see Fig. 1b), which is supposed to be sufficiently large so that reflected waves on the bottom boundary of the model may be neglected, a fixed boundary was imposed. The symmetry conditions also impose that $u_x = 0$ at the lateral surfaces ($x$ = -1 mm and $x$ = 1 mm for sinusoidal profiles; $x = \frac{-L}{2}$ and $x = \frac{L}{2}$ for original profiles).

### B. Finite element simulation

The system of dynamic equations was solved in the time domain using a finite element software (COMSOL Multiphysics, Stockholm, Sweden). The implicit direct time integration generalized-α scheme was used to calculate the transient solution. The elements size was chosen equal to $\lambda_{min}/10$, where $\lambda_{min}$ corresponds to the shortest wavelength in the simulation subdomain. The implant and bone subdomains were meshed by structured quadrangular quadratic elements and the soft tissue subdomain was meshed with triangular quadratic elements. The time step was chosen using the stability Courant-Friedrichs-Lewy (CFL) condition $\Delta t \leq \alpha \min(h_e/c)$ where $\alpha = 1/\sqrt{2}$, $h_e$ is the elements size and $c$ is the velocity in the considered subdomain. For simulations presented here, the time step is set at $\Delta t = 4 \times 10^{-10}$ s. The duration of the simulations was equal to 1.25 µs.

### C. Signal processing

The reflection coefficient was determined for each simulated configuration. To do so, the signal corresponding to the displacement along the direction of propagation was averaged along a horizontal line located at y = $H_{Ti}/2$. The signal corresponding to the averaged incident (respectively reflected) signal was noted $s_i(t)$ (respectively $s_r(t)$). The reflection coefficient in amplitude was determined following:

$$r = A_r/A_i, \qquad (2)$$

where $A_i$ and $A_r$ are respectively the maximum amplitudes of the envelopes of $s_i(t)$ and $s_r(t)$ obtained using the modulus of their respective Hilbert's transform.

### D. Construction of the bone-implant interface

The implant surface roughness was obtained from twenty-one 5 mm diameter coin-shaped implants similar to the ones employed in (Vayron *et al.*, 2012a) and made of medical grade 5 titanium alloy (Ti-6Al-4V). Twelve implants had their surface modified by laser impacts (Faeda *et al.*, 2009; Shah *et al.*, 2016), and nine implants were produced using the EOS supplied Ti-6Al-4V ELE powder and an EOSINT M280 LPBF system (EOS GmbH, Munich, Germany) equipped with a 400 W ytterbium fiber laser. Different levels of surface roughness were obtained by varying the implant orientation in respect to the building platform from 0º (parallel to the platform) to 135º. The roughness profiles of each implant were obtained using a contact profilometer (VEECO Dektak 150) on a 2 mm long line for each sample. The output of each measurement was given by the variation of the surface altitude as a function of the position with a sampling distance of 0.2 µm. The stylus tip had a radius of 2 µm and a force corresponding to 1.00 mg was applied in order to ensure contact between the stylus and the surface at all times.

Different parameters were used to describe the roughness profiles: the average mean roughness $R_a$, the mean spacing between irregularities $S_m$, the maximum profile peak height $R_p$, and the maximum profile valley depth $R_v$. A fifth parameter *s* was introduced to describe the degree of similarity of the original profile with a sinusoidal function and was defined as *s* =

$R_p + R_v - \pi.R_a$. A value of $s = 0$ corresponds to a sinusoidal profile, while higher values of $s$ suggest large irregularities in the roughness profile.

Each surface roughness profile was modified in order to determine the sensitivity of the ultrasound response of the interface on the spatial frequency content of the surface profile. To do so, a low-pass Hamming filter was employed with different cut-off lengths $L_c$ comprised between 2.5 and 500 µm. Note that for the sake of consistency, when comparing the original profile with the filtered ones, the origin for the definition of $W$ always corresponds to the highest position of the original profile.

For each filtered surface profile, the ultrasonic response of the BII was simulated for 13 values of soft tissue thickness $W$ given by: $\{ W = W_k = \frac{k\,W_M}{12}, k \in [0,12]\}$, where $W_M$ depends on the roughness profile and represents a soft tissue thickness value above which the reflection coefficient $r$ of the BII does not vary. Values of $W_M$ were comprised between 55 and 160 µm according to the surface roughness. The range of variation of $W$ was chosen in order to obtain a correct approximation of the ultrasonic response of the BII for various stages of the osseointegration process, including a fully osseointegrated interface ($W = 0$) and a fully debonded interface ($W = W_M$). For each value of $W_k$, the value of the simulated reflection coefficient $r_o(k)$ obtained with the original profile was compared with the values of the reflection coefficient $r_f(L_c, k)$ obtained with a filtered profile (with a cut-off length $L_c$) and the same soft tissue thickness of $W_k$. The maximum difference $D(L_c)$, when varying $W$ between the reflection coefficients obtained with the original profile and the corresponding filtered profile (with a cut-off length equal to $L_c$), was defined as:

$$D(Lc) = \max_{k \in [0;12]} (|r_o(k) - r_f(Lc, k)|) \qquad (3)$$

### E. Determination of the optimal equivalent sinusoidal profile

For each profile corresponding to the samples described in subsection II.D, an optimization method was developed in order to determine the equivalent sinusoidal profile (with roughness parameters ($h_{eq}$, $L_{eq}$)) leading to an ultrasonic response that best matches the ones obtained with the original roughness profile. This optimization method is illustrated in Fig. 2 and described below.

The comparison between the reflection coefficient obtained with the original and the sinusoidal equivalent profiles was realized based on the difference of the reflection coefficients obtained for 13 different values of $W$ defined by: $\{ W = W_k = \frac{k\,W_M}{12}, k \in [0,12]\}$, similarly as what was done in subsection II.D. For each original profile, the reflection coefficient $r_o(k)$ was determined for $W = W_k, k \in [0,12]$. The values of the reflection coefficient $r_o(k)$ obtained with the original profile were compared with the values of the reflection coefficient $r_{sin}(h, L, k)$ obtained with a sinusoidal profile having for roughness parameters: ($h$, $L$)) and with the soft tissue thickness equal to $W_k$. A cost function $e(h,L)$ was defined in order to assess the difference between the ultrasound response of the original and each sinusoidal profile (with roughness parameters (*h*, *L*) following:

$$e(h, L) = \sum_{k=0}^{12} \frac{|r_o(k) - r_{sin}(h, L, k)|}{13} \quad (4)$$

An optimization procedure based on a conjugate gradient method (Nazareth, 2009) was carried out in order to determine the optimal values of the roughness parameters (*h*, *L*) minimizing the cost function *e*. The algorithm was initiated for $h = \pi R_a$ and $L = S_m/2$ because these parameters correspond to the values that would be obtained if the original profile was sinusoidal.

Two convergence criteria that must both be achieved to consider the process as converged were set as:

$$\sqrt{(h_i - h_{i-1})^2 + (L_i - L_{i-1})^2} < 0.02 \ \mu m \quad (5)$$

$$e_i - e_{i-1} < 10^{-5} \quad (6)$$

where *i* is the number of iterations performed.

## III. Results

### A. Influence of roughness on the ultrasonic response

Table II shows the roughness parameters of the 21 original profiles. Implants with laser-modified surfaces have significantly lower roughness amplitude $R_a$ and higher values of $S_m$ compared to 3D-printed implants.

Figure 3 shows the rf signals with their envelopes corresponding to the simulated ultrasonic waves recorded at $y = H_{Ti}/2$ and averaged over *x* for an implant surface with Ra = 24.2 μm in the cases of fully-bonded (*W* = 0) and fully debonded ($W = W_M$) interfaces. Note that the results did not significantly vary when the convergence criteria given in Eqs. 5-6 were decreased, which constitutes a validation of the approach.

Figure 4 shows the variation of the reflection coefficient $r_o(k)$ obtained for different roughness profiles as a function of the soft tissue thickness $W_k$, $k \in [0,12]$. Figure 4a shows the results obtained for the surface profiles corresponding to implants with laser-modified surface, which have a relatively low surface roughness. The results obtained with the different surface profiles are qualitatively similar. The values of $r_o$ first increase as a function of *W* from 0.54 to a maximum value equal to around 0.92. Then, $r_o$ slightly decreases as a function of *W* and tends towards 0.88 for all the profiles considered. However, an increase in $r_o$ occurs for smaller values of *W* when considering surfaces with lower roughness. Similarly, the maximum value of $r_o$ is reached for lower values of *W* when considering surfaces with lower roughness. The maximum peak height $R_p$ of the surface profile seems to have a higher influence on the variation of $r_o$ than the average roughness amplitude $R_a$, because the roughness profiles with similar $R_p$ lead to approximately similar ultrasonic responses. Note that the values of the reflection coefficient $r_o(0)$ obtained for $W_0 = 0$ (respectively $r_o(12)$ for $W_{12} = 100$ μm) correspond to the analytical values obtained for a planar bone-implant interface (respectively soft tissue-implant interface) and are weakly affected by the profile roughness.

Figure 4b shows the results obtained for 3D-printed implants, which corresponds to relatively important surface roughness. The reflection coefficient $r_o$ is shown to first increase as a function of $W$ and then to tend towards constant values above approximatively $W = R_v + R_p$ for all the profiles considered. Again, the increase of $r_o$ occurs for lower values of $W$ when considering surface profiles with lower values of $R_p$. Moreover, the values of $r_o$ obtained for $W_o=0$ and for $W_{12}=200$ μm increase as a function of $R_a$, which constitutes a different situation compared to the case of implants with laser-modified surfaces (see Fig. 4a) and will be discussed in subsection IV.B.

### B.  Effect of low-pass filtering the surface profile

Figure 5a (respectively 5b) show the different profiles obtained after application of the low-pass filters with different values of the cut-off length $L_c$ varying between 10 and 125 μm to the original roughness profile with $R_a = 5.83$ μm (respectively $R_a = 18.2$ μm). The original profile is also shown. Figure 5c (respectively 5d) shows the variation of the reflection coefficient $r$ as a function of $W$ for the different surface profiles shown in Fig. 5a (respectively 5b). Figure 5c (corresponding to an original profile with $R_a = 5.83$ μm) indicates that the variation of $r$ as a function of $W$ is approximately similar for all filtered profiles. Namely, $r$ first increases as a function of $W$ from around 0.54 to reach a maximum value equal to around 0.92. Then, $r$ decreases as a function of $W$ and tends towards around 0.88. However, Fig. 5d (corresponding to $R_a = 18.2$ μm for the original profile) indicates that the variation of $r$ as a function of $W$ varies according to the filtered profiles. Again, $r$ first increases as a function of $W$, but reaches a maximum value that increases as a function of $L_c$. Moreover, for a given value of $W$, $r$ is shown to increase as a function of $L_c$, which may be explained by a progressive decrease of scattering phenomena when the profile is filtered in Fig. 5d, whereas the initial roughness was not sufficient to cause scattering effects in the case of the laser modified surface (see Fig. 5c).

Figure 6a shows the variation of the difference $D$ between the reflection coefficient of the filtered profiles and of the corresponding original profiles as a function of $L_c$. The results are shown for three implants with laser-modified surfaces and three implants with 3D-printed surfaces. For all profiles, $D$ increases as a function of $L_c$ and then reaches a constant value when $L_c$ tends towards infinity.

Figure 6b shows the variation of the average roughness $R_a$ of the profiles as a function of $L_c$. For implants with laser-modified surfaces, $R_a$ decreases significantly for low values of $L_c$ and is close to 0 for $L_c > 250$ μm. However, for 3D-printed implants, which had a higher values of $R_a$ and $S_m$, thus implying more high frequency components, $R_a$ continues to decrease significantly for $L_c > 500$ μm. Consequently, D converges more quickly towards its final value for implants with laser-modified surfaces compared to 3D-printed implants, as shown in Fig. 6a.

### C. Optimal equivalent sinusoidal profile

Figure 7 shows three original roughness profiles and their respective equivalent sinusoidal profiles determined using the optimization procedure described in subsection II.E.

Figure 8 shows the variation of the reflection coefficient $r$ as a function of $W$ for the same roughness profiles as the ones shown in Fig. 7 and for their respective equivalent sinusoidal profiles. The values of $r$ were determined for each value of $W$ for which the cost function $e$ was evaluated (see Eq. 4). The results show that the behavior of $r$ is qualitatively the same for the original and for the equivalent sinusoidal profile. However, the minimum value of the cost function is shown to increase as a function of $R_a$.

Figure 9a (respectively 9b) shows that $h_{eq}$ increases as a function of $R_a$ (respectively $R_p$) for all original profiles. Second-order polynomial regressions can approximate the dependence of $h_{eq}$ as a function of both $R_a$ and $R_p$. However, Spearman's tests indicated a better correlation between $h$ and $R_p$ ($r_S$= 0.997) compared to the correlation between $h$ and $R_a$ ($r_S$= 0.970), which will be discussed in subsection IV.C. Moreover, Fig. 9a shows that the amplitude $h_{eq}$ of each equivalent sinusoidal profile is always comprised between π.$Ra$, which would be the value of $h$ if the original profile was sinusoidal, and $R_p + R_v$, which represents the maximum amplitude of the original profile.

Figure 10a (respectively 10b) shows that $L_{eq}$ increases as a function of $R_a$ (respectively $S_m$). For all implants with laser-modified surfaces, $L_{eq}$ stays relatively constant as 83% of the values of $L_{eq}$ are comprised between 54 and 58 μm. For 3D-printed implants, the values of $L_{eq}$ are significantly higher and depend on the roughness of the original profile. Spearman's tests indicate significant correlations between $L_{eq}$ and both $R_a$ ($r_S$= 0.843) and $S_m$ ($r_S$= 0.833), which will be discussed in subsection IV.C.

Figure 11a (respectively 11b) illustrates that the minimum value of the cost function $e_{min}$ increases as a function of $R_a$ (respectively of $s = R_p + R_v - \pi.R_a$). Spearman's tests indicate a significant correlation between $L_{eq}$ and $R_a$ ($r_S$= 0.848) and a stronger one between $L_{eq}$ and $s$ ($r_S$= 0.911), which describes the similarity of the original profile with a sinusoidal variation.

## IV. Discussion

### A. Originality and comparison with literature

The originality of this study is to consider a realistic description of the bone-implant interface and to analyze the effect of the different roughness parameters and of osseointegration phenomena on the ultrasonic response of the BII. Previous numerical studies (Vayron *et al.*, 2015; 2016) have investigated the variation of the ultrasonic response of the BII during the osseointegration process, which was modeled by a variation of bone properties around the implant. In these two previous papers, a fully bonded BII and an absence of osseointegration were the two cases considered. The effect of the microscopic implant roughness was not accounted for. (Heriveaux *et al.*, 2018) is the only study investigating the impact of microscopic implant roughness on the ultrasonic response of the BII but a sinusoidal profile was then considered. The variation of *r* as a function of *W* obtained in (Heriveaux *et al.*, 2018) in the case of a microscopic roughness is in qualitative agreement with the results of the present paper (see Fig. 4a), which justifies the comparison between both models developed in sections II.E and III.C.

Different experimental studies have also evidenced the effect of osseointegration phenomena on the ultrasonic response of the BII. In particular, the effect of healing time on the ratio between the amplitudes of the echo of the BII and of the water-implant interface was studied in (Vayron *et al.*, 2012a) using implants with an average roughness of $R_a = 1.9$ μm, which is of the same order of magnitude as the implants considered in this study (see Fig 4a and Table II). Mathieu et al. (2012) found a decrease of the apparent reflection coefficient of 7.8% between 7 and 13 weeks of healing time, which corresponds to an increase of the BIC from 27 to 69%. The model considered herein predicts that an increase of the BIC from 27 to 69% should result in a decrease of *r* by 7.3% in the case $R_a = 1.52$ μm, and by 10.7%, in the case $R_a = 2.52$ μm, which is relatively close to the experimental results. However, some discrepancies could explain the differences between experimental results and numerical predictions. First, the present study does not consider the changes of the bone material properties, which are known to occur during healing (Mathieu *et al.*, 2011b; Vayron *et al.*, 2012b; Vayron *et al.*, 2014b) and which induce a concurrent increase of the reflection coefficient as a function of healing time (Vayron *et al.*, 2016). Second, in the experimental configuration, the ultrasonic wave is not fully planar due to the use of a focused immersed transducer, which has not been considered in the present study. Despite these limitations, a good agreement is obtained between numerical and experimental results.

Another set of studies (Vayron *et al.*, 2014c; Vayron *et al.*, 2018a) have investigated the variation of the 10 MHz echographic response of a dental implant using a dedicated ultrasound device (Vayron *et al.*, 2014c). These studies showed that the amplitude of the echographic response of a dental implant decreases as a function of healing time, which is in qualitative agreement with the present study. However, a quantitative comparison is difficult due to the complex geometry of dental implants.

The averaged experimental error $\varepsilon$ on the determination of the reflection coefficient by ultrasonic methods was found equal to $3.10^{-2}$ (Vayron *et al.*, 2012a). Therefore, considering results presented in Fig. 6, the difference between real and filtered profiles would be detected for cut-off lengths between 45 and 120 μm, except for the case of the profile with $R_a = 0.9$ μm. For this last profile, since the roughness is already low, the difference of ultrasonic response with a perfectly smooth implant would not be detectable. Moreover, the minimum value of the cost-function $e_{min}$ corresponding to an averaged difference of *r* obtained between the original profile and its equivalent sinusoidal profile was comprised between $2.2.10^{-3}$ and $3.2.10^{-2}$ (see Table II), which is lower or of the same order of magnitude compared to the experimental error $\varepsilon$, and constitutes a validation of the approach developed in section II.E.

## B. Influence of the roughness

The results shown in Fig. 4 illustrate that the reflection coefficient of the BII depends on the surface roughness of the implant. In particular, two distinct behaviors may be observed depending whether the implants have a relatively low (implants with laser-modified surfaces, see Fig. 4a) or high (3D-printed implants, see Fig. 4b) surface roughness.

In the cases of a fully-bonded interface ($W=0$) and of no osseointegration ($W=100$ μm), Fig. 4a shows that $r_o$ is approximately constant for implants with low surface roughness, while for implants with higher surface roughness, $r_o$ decreases as a function of $R_a$, as shown in Fig. 4b. The decrease of $r_o$ as a function of $R_a$ may be explained by scattering effects of the wave on the BII, which increases with the roughness amplitude.

Figure 4a shows that for implants with laser-modified surfaces, $r_o$ reaches a local maximum for a value of $W$ comprised between 30 and 60 μm depending on the surface roughness. This result may be explained by constructive interferences of the echoes of the soft tissue - bone interface and of the implant - soft tissue interface, as already described in (Heriveaux *et al.*, 2018). When the value of $W$ is sufficiently high so that these interferences disappear, $r_o$ finally decreases to reach a final value of around 0.88. To a lesser extent, the effects of these interferences may also be observed for 3D-printed implants (see Fig. 4b), as $r_o$ also reaches a local maximum. For this latter group of implants, $r_o$ eventually converges for $W \sim R_v + R_p$ towards a value comprised between 0.72 and 0.85 depending on the roughness profile, because when $W > R_v + R_p$, BIC = 0 and no bone is in contact with the implant surface.

For all implants, $r_o$ starts to increase for lower values of $W$ when considering lower values of $R_p$. Figures 4a and 4b illustrate that the value of soft tissue thickness $W_{R=0.6}$ for which $r_o$ reaches a value of 0.6 increases as a function of $R_p$. An explanation of this behavior is provided by the geometrical definition of $R_p$, which induce that $R_p$ is closely related to the value of soft tissue thickness $W_{50}$ corresponding to a BIC value of 50%. Therefore, the BIC value corresponding to a given value of soft tissue thickness $W$ tends to increase as a function of $R_p$. Since $r_o$ is also an increasing function of the BIC, the aforementioned results explain that $W_{R=0.6}$ is an increasing function of $R_p$.

## C. Equivalence of the sinusoidal model

Figure 9 shows that the behavior of $h_{eq}$ is more closely related to variations of $R_p$ than to variations of $R_a$, which may be explained by the interpretation given in subsection IV.B. $R_p$ is shown to strongly influence the value of soft tissue thickness $W$ at which $r_o$ starts to increase and more generally the behavior of $r_o$ as a function of $W$. These results explain the important effect of $R_p$ on the value of $h_{eq}$ because $r$ should have the same dependence on $W$ for the original and for the equivalent sinusoidal profiles in order to minimize the cost function $e_{min}$. Nevertheless, $R_a$ and $R_p$ being interdependent, a significant correlation between $h_{eq}$ and $R_a$ was also obtained in Fig. 9a.

As shown in Fig. 9b, the variation of $h_{eq}$ as a function of $R_p$ can be well approximated by a second order polynomial variation given by:

$$h_{eq} = 1.94\, R_p - 0.0083\, R_p^2. \qquad (9)$$

Equation (9) may be used in the future to initialize the optimization process described in Fig. 2 in order to achieve a faster convergence. Moreover, Eq. (9) may be explained as follows. Perfectly sinusoidal profiles would lead to the relation: $h_{eq} = 2\, R_p$. For low values of $R_p$, the original profiles also have a low value of $s$ (see Table II), which explains that $h_{eq} \approx 2\, R_p$ when $R_p$ tends towards 0. When $R_p$ further increases, the original profiles become more different compared to sinusoidal variations, which explains the second term ($-0.0083\, R_p^2$).

Figure 10 shows a significant correlation between $L_{eq}$ and $R_a$, especially for 3D-printed implants. However, a better correlation would have been expected between $L_{eq}$ and $S_m$ (Fig. 10a) than between $L_{eq}$ and $R_a$ (Fig. 10b), which is not the case. It may be explained by the fact that $S_m$ strongly depends on local peaks and may therefore not be an accurate indicator of the periodicity of the roughness profiles.

### D. Limitations

This study has several limitations. First, only normal incidence of the ultrasonic wave was considered as it corresponds to an experimental situation of interest (Vayron *et al.*, 2012a; Vayron *et al.*, 2014c). Second, adhesion phenomena at the BII (Vayron *et al.*, 2012a), which may cause a non-linear ultrasonic response (Biwa *et al.*, 2004), were not considered herein. Third, the variation of the periprosthetic bone geometrical properties was modeled by a bone level given by the parameter $W$ and actual bone geometry around the implant surface is likely to be more complex. Note that typical BIC values are comprised between 30 and 80% (Scarano *et al.*, 2006; Vayron *et al.*, 2012a; Pontes *et al.*, 2014; Vayron *et al.*, 2014c). Therefore, fully-bonded interfaces are not likely to occur *in vivo*. Moreover, bone properties are known to vary during osseointegration (Mathieu *et al.*, 2011b; Vayron *et al.*, 2012b; Vayron *et al.*, 2014b), which was not taken into account. Fourth, bone tissue was modeled as an elastic, homogeneous and isotropic material, similarly to what was done in some previous studies (Haïat *et al.*, 2009; Mathieu *et al.*, 2011a; Vayron *et al.*, 2015; 2016), whereas real bone tissue is known to be a strongly dispersive medium (Naili *et al.*, 2008; Haiat and Naili, 2011). Moreover, although mature bone tissue is known to be anisotropic (Haïat *et al.*, 2009; Sansalone *et al.*, 2012), the anisotropic behavior of newly formed bone tissue remains unknown (Mathieu *et al.*, 2011b; Vayron *et al.*, 2012a). Fifth, the study only focused on a frequency of 10 MHz because it corresponds to a frequency used for characterization purposes (Vayron *et al.*, 2018a). However, LIPUS used for stimulation purposes would have lower frequencies (Dimitriou and Babis, 2007), which was not investigated herein. Sixth, we only consider the first reflection of the ultrasonic wave on the BII, similarly as what was done in (Mathieu *et al.*, 2012), because it constitutes a simple approach to determine the effect of variations of the properties of the BII on its ultrasonic response. Last, two-dimensional modeling of the BII was considered and the 3-D results may be different. Future works should focus on a 3-D description of the interface and on improving the modeling of osseointegration phenomena to derive a more realistic description of the interaction between ultrasound and the BII.

**Acknowledgements**

This project has received funding from the European Research Council (ERC) under the European Union's Horizon 2020 research and innovation program (grant agreement No 682001, project ERC Consolidator Grant 2015 *BoneImplant*).

**References**

Biwa, S., Nakajima, S., and Ohno, N. (**2004**). "On the Acoustic Nonlinearity of Solid-Solid Contact With Pressure-Dependent Interface Stiffness," Journal of Applied Mechanics **71**, 508-515.

Claes, L., and Willie, B. (**2007**). "The enhancement of bone regeneration by ultrasound," Prog Biophys Mol Biol **93**, 384-398.

de Almeida, M. S., Maciel, C. D., and Pereira, J. C. (**2007**). *Proposal for an Ultrasonic Tool to Monitor the Osseointegration of Dental Implants* (Sensors (Basel). 2007 Jul;7(7):1224-37.).

Dimitriou, R., and Babis, G. (**2007**). "Biomaterial osseointegration enhancement with biophysical stimulation," J Musculoskelet Neuronal Interact **7**, 257-263.

Duarte, L. R. (**1983**). "The stimulation of bone growth by ultrasound," Archives of orthopaedic and traumatic surgery **101**, 153-159.

Faeda, R. S., Tavares, H. S., Sartori, R., Guastaldi, A. C., and Marcantonio, E., Jr. (**2009**). "Evaluation of titanium implants with surface modification by laser beam. Biomechanical study in rabbit tibias," Brazilian oral research **23**, 137-143.

Franchi, M., Bacchelli, B., Giavaresi, G., De Pasquale, V., Martini, D., Fini, M., Giardino, R., and Ruggeri, A. (**2007**). "Influence of Different Implant Surfaces on Peri-Implant Osteogenesis: Histomorphometric Analysis in Sheep," Journal of Periodontology **78**, 879-888.

Georgiou, A. P., and Cunningham, J. L. (**2001**). "Accurate diagnosis of hip prosthesis loosening using a vibrational technique," Clinical biomechanics (Bristol, Avon) **16**, 315-323.

Gill, A., and Shellock, F. G. (**2012**). "Assessment of MRI issues at 3-Tesla for metallic surgical implants: findings applied to 61 additional skin closure staples and vessel ligation clips," Journal of cardiovascular magnetic resonance : official journal of the Society for Cardiovascular Magnetic Resonance **14**, 3.

Haiat, G., and Naili, S. (**2011**). "Independent scattering model and velocity dispersion in trabecular bone: comparison with a multiple scattering model," Biomechanics and Modeling in Mechanobiology **10**, 95-108.

Haïat, G., Naili, S., Grimal, Q., Talmant, M., Desceliers, C., and Soize, C. (**2009**). "Influence of a gradient of material properties on ultrasonic wave propagation in cortical bone: Application to axial transmission," The Journal of the Acoustical Society of America **125**, 4043-4052.

Heller, A. L., and Heller, R. L. (**1996**). "Clinical evaluations of a porous-surfaced endosseous implant system," The Journal of oral implantology **22**, 240-246.

Heriveaux, Y., Nguyen, V. H., and Haiat, G. (**2018**). "Reflection of an ultrasonic wave on the bone-implant interface: A numerical study of the effect of the multiscale roughness," The Journal of the Acoustical Society of America **144**, 488.

Khan, S. N., Ramachandran, M., Senthil Kumar, S., Krishnan, V., and Sundaram, R. (**2012**). "Osseointegration and more–A review of literature," Indian Journal of Dentistry **3**, 72-76.

Madsen, E. L., Sathoff, H. J., and Zagzebski, J. A. (**1983**). "Ultrasonic shear wave properties of soft tissues and tissuelike materials," The Journal of the Acoustical Society of America **74**, 1346-1355.

Mathieu, V., Anagnostou, F., Soffer, E., and Haiat, G. (**2011a**). "Numerical simulation of ultrasonic wave propagation for the evaluation of dental implant biomechanical stability," J. Acoust. Soc. Am. **129**, 4062-4072.

Mathieu, V., Fukui, K., Matsukawa, M., Kawabe, M., Vayron, R., Soffer, E., Anagnostou, F., and Haiat, G. (**2011b**). "Micro-Brillouin scattering measurements in mature and newly formed bone tissue surrounding an implant," J Biomech Eng **133**, 021006.

Mathieu, V., Michel, A., Flouzat Lachaniette, C. H., Poignard, A., Hernigou, P., Allain, J., and Haiat, G. (**2013**). "Variation of the impact duration during the in vitro insertion of acetabular cup implants," Medical engineering & physics **35**, 1558-1563.

Mathieu, V., Vayron, R., Richard, G., Lambert, G., Naili, S., Meningaud, J. P., and Haiat, G. (**2014**). "Biomechanical determinants of the stability of dental implants: influence of the bone-implant interface properties," J Biomech **47**, 3-13.

Mathieu, V., Vayron, R., Soffer, E., Anagnostou, F., and Haiat, G. (**2012**). "Influence of healing time on the ultrasonic response of the bone-implant interface," Ultrasound Med Biol **38**, 611-618.

Meredith, N., Alleyne, D., and Cawley, P. (**1996**). "Quantitative determination of the stability of the implant-tissue interface using resonance frequency analysis," Clinical oral implants research **7**, 261-267.

Michel, A., Bosc, R., Meningaud, J.-P., Hernigou, P., and Haiat, G. (**2016**). "Assessing the Acetabular Cup Implant Primary Stability by Impact Analyses: A Cadaveric Study," PLOS ONE **11**, e0166778.

Moerman, A., Zadpoor, A. A., Oostlander, A., Schoeman, M., Rahnamay Moshtagh, P., Pouran, B., and Valstar, E. (**2016**). "Structural and mechanical characterisation of the peri-prosthetic tissue surrounding loosened hip prostheses. An explorative study," Journal of the mechanical behavior of biomedical materials **62**, 456-467.

Naili, S., Padilla, F., and Laugier, P. (**2008**). "Fast wave propagation in trabecular bone: numerical study of the influence of porosity and structural anisotropy," Journal of the Acoustical Society of America **123**, 1694-1705.

Nakanishi, Y., Wang, P.-L., Ochi, M., Nakanishi, K., and Matsubara, H. (**2011**). "Low-intensity Pulsed Ultrasound Stimulation Significantly Enhances the Promotion of Bone Formation Around Dental Implants," Journal of Hard Tissue Biology **20**, 139-146.

Nazareth, J. L. (**2009**). "Conjugate gradient method," Wiley Interdisciplinary Reviews: Computational Statistics **1**, 348-353.

Njeh, C. F., Hans, D., Wu, C., Kantorovich, E., Sister, M., Fuerst, T., and Genant, H. K. (**1999**). "An in vitro investigation of the dependence on sample thickness of the speed of sound along the specimen," Medical engineering & physics **21**, 651-659.

Padilla, F., Peyrin, F., Laugier, P., and Haiat, G. (**2007**). *Variation of Ultrasonic Parameters With Microstructure and Material Properties of Trabecular Bone: A 3D Model Simulation*.

Padilla, F., Puts, R., Vico, L., and Raum, K. (**2014**). "Stimulation of bone repair with ultrasound: A review of the possible mechanic effects," Ultrasonics **54**, 1125-1145.

Pastrav, L. C., Jaecques, S. V., Jonkers, I., Perre, G. V., and Mulier, M. (**2009**). "In vivo evaluation of a vibration analysis technique for the per-operative monitoring of the fixation of hip prostheses," Journal of orthopaedic surgery and research **4**, 10.

Pattijn, V., Jaecques, S. V. N., De Smet, E., Muraru, L., Van Lierde, C., Van der Perre, G., Naert, I., and Vander Sloten, J. (**2007**). "Resonance frequency analysis of implants in the guinea pig model: Influence of boundary conditions and orientation of the transducer," Medical engineering & physics **29**, 182-190.

Pattijn, V., Van Lierde, C., Van der Perre, G., Naert, I., and Vander Sloten, J. (**2006**). "The resonance frequencies and mode shapes of dental implants: Rigid body behaviour versus bending behaviour. A numerical approach," Journal of Biomechanics **39**, 939-947.

Pilliar, R. M., Lee, J. M., and Maniatopoulos, C. (**1986**). "Observations on the effect of movement on bone ingrowth into porous-surfaced implants," Clinical orthopaedics and related research, 108-113.

Pontes, A. E. F., Ribeiro, F. S., Iezzi, G., Pires, J. R., Zuza, E. P., Piattelli, A., and Marcantonio Junior, E. (**2014**). "Bone-Implant Contact around Crestal and Subcrestal Dental Implants Submitted to Immediate and Conventional Loading," The Scientific World Journal **2014**, 606947.

Sansalone, V., Bousson, V., Naili, S., Bergot, C., Peyrin, F., Laredo, J. D., and Haïat, G. (**2012**). "Anatomical distribution of the degree of mineralization of bone tissue in human femoral neck: Impact on biomechanical properties," Bone **50**, 876-884.

Sarvazyan, A. P., Urban, M. W., and Greenleaf, J. F. (**2013**). "Acoustic waves in medical imaging and diagnostics," Ultrasound Med Biol **39**, 1133-1146.

Scarano, A., Degidi, M., Iezzi, G., Petrone, G., and Piattelli, A. (**2006**). "Correlation between implant stability quotient and bone-implant contact: a retrospective histological and histomorphometrical study of seven titanium implants retrieved from humans," Clinical implant dentistry and related research **8**, 218-222.

Schulte, W., d'Hoedt, B., Lukas, D., Muhlbradt, L., Scholz, F., Bretschi, J., Frey, D., Gudat, H., Konig, M., Markl, M., and et al. (**1983**). "[Periotest--a new measurement process for periodontal function]," Zahnarztliche Mitteilungen **73**, 1229-1230, 1233-1226, 1239-1240.

Shah, F. A., Johansson, M. L., Omar, O., Simonsson, H., Palmquist, A., and Thomsen, P. (**2016**). "Laser-Modified Surface Enhances Osseointegration and Biomechanical Anchorage of Commercially Pure Titanium Implants for Bone-Anchored Hearing Systems," PLOS ONE **11**, e0157504.

Shalabi, M., Wolke, J., Cuijpers, V., and A Jansen, J. (**2007**). *Evaluation of bone response to titanium-coated polymethyl methacrylate resin (PMMA) implants by X-ray tomography*.

Tanzer, M., Kantor, S., and Bobyn, J. (**2001**). "Enhancement of bone growth into porous intramedullary implants using non-invasive low intensity ultrasound," Journal of orthopaedic research : official publication of the Orthopaedic Research Society **19**, 195-199.

Van Scotter, D. E., and Wilson, C. J. (**1991**). "The Periotest method for determining implant success," The Journal of oral implantology **17**, 410-413.

Vayron, R., Barthel, E., Mathieu, V., Soffer, E., Anagnostou, F., and Haiat, G. (**2012a**). "Nanoindentation Measurements of Biomechanical Properties in Mature and Newly Formed Bone Tissue Surrounding an Implant," Journal of Biomechanical Engineering **134**, 021007-021007-021006.

Vayron, R., Barthel, E., Mathieu, V., Soffer, E., Anagnostou, F., and Haiat, G. (**2012b**). "Nanoindentation measurements of biomechanical properties in mature and newly formed bone tissue surrounding an implant," J Biomech Eng **134**, 021007.

Vayron, R., Karasinski, P., Mathieu, V., Michel, A., Loriot, D., Richard, G., Lambert, G., and Haiat, G. (**2013**). "Variation of the ultrasonic response of a dental implant embedded in tricalcium silicate-based cement under cyclic loading," Journal of Biomechanics **46**, 1162-1168.

Vayron, R., Mathieu, V., Michel, A., and Haïat, G. (**2014a**). "Assessment of In Vitro Dental Implant Primary Stability Using an Ultrasonic Method," Ultrasound in Medicine & Biology **40**, 2885-2894.

Vayron, R., Matsukawa, M., Tsubota, R., Mathieu, V., Barthel, E., and Haiat, G. (**2014b**). "Evolution of bone biomechanical properties at the micrometer scale around titanium implant as a function of healing time," Physics in medicine and biology **59**, 1389-1406.

Vayron, R., Nguyen, V.-H., Bosc, R., Naili, S., and Haïat, G. (**2015**). "Finite element simulation of ultrasonic wave propagation in a dental implant for biomechanical stability assessment," Biomech Model Mechanobiol **14**, 1021-1032.

Vayron, R., Nguyen, V.-H., Bosc, R., Naili, S., and Haïat, G. (**2016**). "Assessment of the biomechanical stability of a dental implant with quantitative ultrasound: A three-dimensional finite element study," The Journal of the Acoustical Society of America **139**, 773-780.

Vayron, R., Nguyen, V.-H., Lecuelle, B., Albini Lomami, H., Meningaud, J.-P., Bosc, R., and Haiat, G. (**2018a**). "Comparison of Resonance Frequency Analysis and of Quantitative Ultrasound to Assess Dental Implant Osseointegration," Sensors (Basel, Switzerland) **18**, 1397.

Vayron, R., Nguyen, V. H., Lecuelle, B., and Haiat, G. (**2018b**). "Evaluation of dental implant stability in bone phantoms: Comparison between a quantitative ultrasound technique and resonance frequency analysis," Clinical implant dentistry and related research.

Vayron, R., Soffer, E., Anagnostou, F., and Haïat, G. (**2014c**). "Ultrasonic evaluation of dental implant osseointegration," Journal of Biomechanics **47**, 3562-3568.

Table I: Material properties used in the numerical simulations.

|  | *Longitudinal velocity $V_L$ (m.s$^{-1}$)* | *Shear velocity $V_S$ (m.s$^{-1}$)* | *Mass density $\rho$ (kg.m$^{-3}$)* |
|---|---|---|---|
| Soft tissue | 1500 | 10 | 1000 |
| Titanium | 5810 [a] | 3115 [a] | 4420 [a] |
| Cortical bone tissue | 4000 | 1800 | 1850 [b] |

[a]See (Pattijn *et al.*, 2006; Pattijn *et al.*, 2007).
[b]See (Njeh *et al.*, 1999; Padilla *et al.*, 2007).

Table II: Roughness parameters of the original profiles

|  | $R_a$ (µm) | $R_p$ (µm) | $R_v$ (µm) | $S_m$ (µm) | $s$ (µm) |
|---|---|---|---|---|---|
| Laser-modified surfaces | 0.898 | 3.28 | 2.94 | 73.9 | 3.41 |
| | 1.29 | 4.74 | 5.46 | 67.6 | 6.14 |
| | 1.44 | 4.74 | 4.24 | 78.0 | 4.46 |
| | 1.52 | 4.85 | 5.27 | 82.5 | 5.34 |
| | 2.52 | 7.40 | 10.9 | 95.0 | 10.4 |
| | 3.13 | 10.5 | 13.3 | 54.2 | 14.0 |
| | 3.92 | 11.1 | 14.6 | 65.1 | 13.5 |
| | 4.13 | 11.5 | 16.4 | 77.6 | 14.9 |
| | 4.93 | 17.1 | 19.5 | 50.6 | 21.1 |
| | 5.77 | 17.0 | 18.9 | 58.1 | 17.8 |
| | 5.83 | 17.9 | 19.6 | 56.4 | 19.3 |
| | 6.94 | 20.4 | 40.0 | 57.2 | 38.6 |
| 3D-printed | 14.0 | 46.8 | 40.3 | 267 | 43.2 |
| | 16.8 | 34.0 | 51.0 | 165 | 32.2 |
| | 18.1 | 58.1 | 33.7 | 171 | 34.8 |
| | 18.2 | 50.5 | 51.1 | 172 | 44.5 |
| | 18.4 | 59.7 | 44.4 | 211 | 46.2 |
| | 19.0 | 53.0 | 45.3 | 175 | 38.6 |
| | 19.7 | 41.4 | 58.1 | 182 | 37.6 |
| | 22.8 | 68.8 | 48.2 | 316 | 45.4 |
| | 24.2 | 69.1 | 50.9 | 206 | 44.0 |

**Figure**

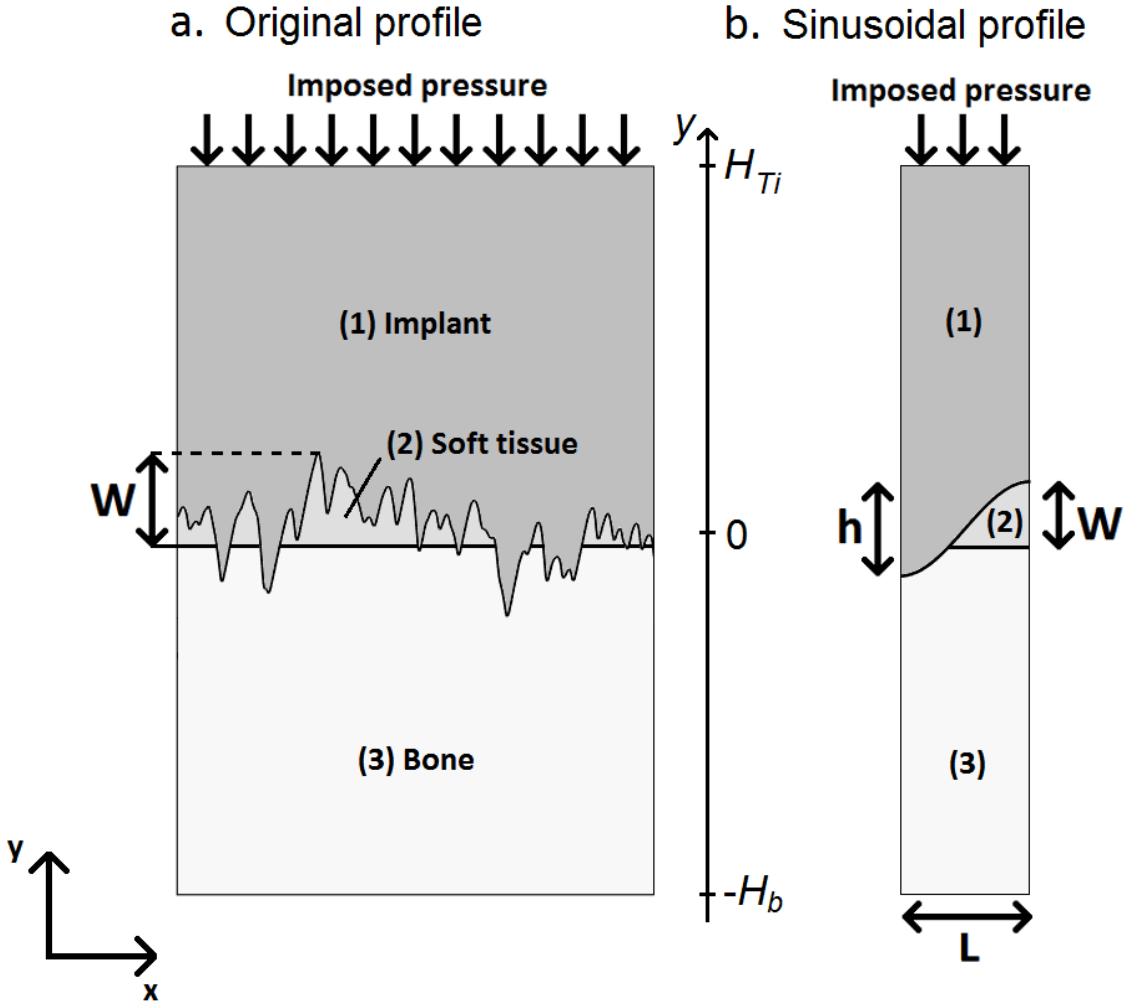

Fig. 1: *Schematic illustrations of the 2-D model used in numerical simulations for (a) an original roughness profile and (b) a sinusoidal roughness profile.*

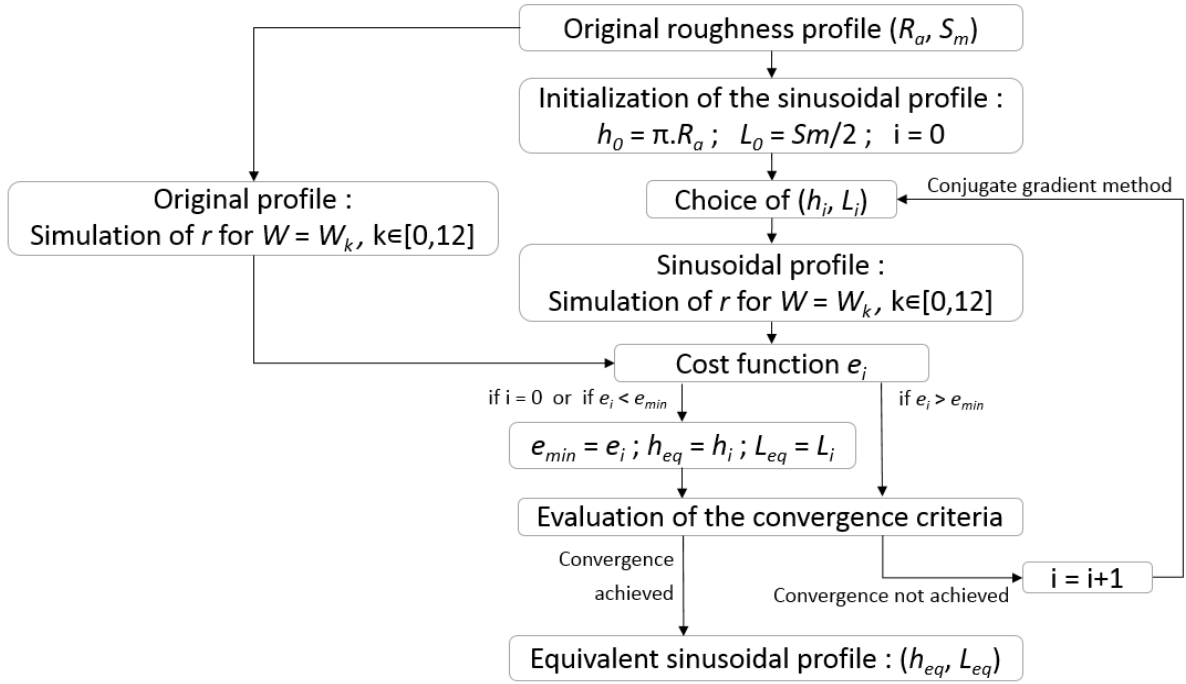

Fig. 2: *Schematical description of the optimization method aiming at determining the equivalent sinusoidal roughness profile (with parameters $h_{eq}$, $L_{eq}$) corresponding to each original roughness profile (with parameters $R_a$, $S_m$).*

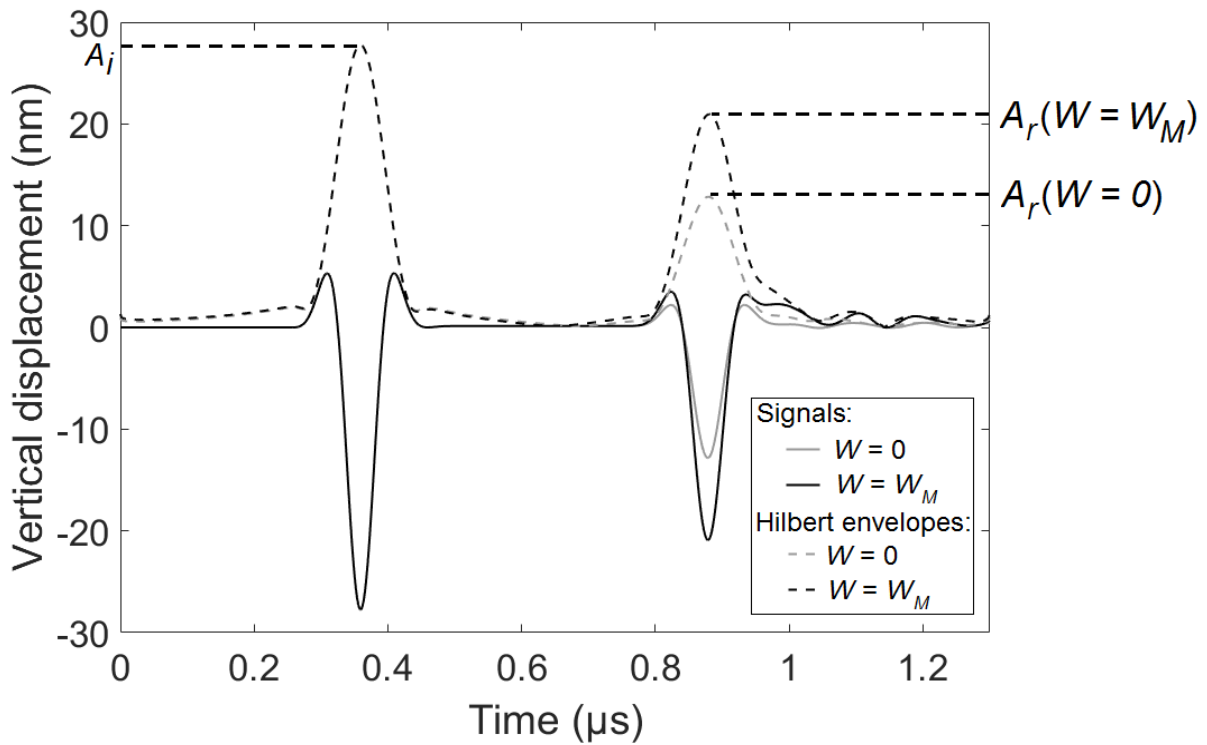

Fig. 3. *Radiofrequency signals (solid lines) with their envelopes (dashed lines) corresponding to the ultrasonic waves recorded at $H_{Ti}/2$ and averaged over x for an implant surface with $R_a$ = 24.2 μm in the cases of fully-bonded (W = 0) and fully debonded (W = $W_M$) interfaces.*

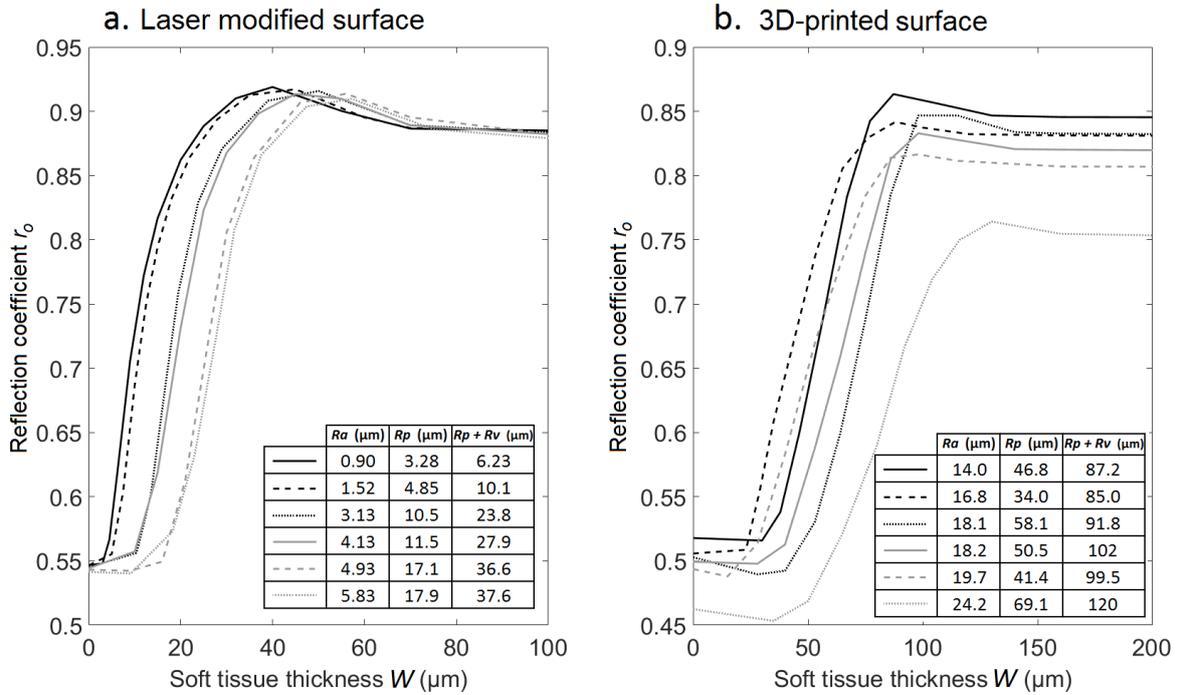

Fig. 4: *Variation of the reflection coefficient $r_o$ of the bone-implant interface as a function of the soft tissue thickness W for (a) six implants with laser-modified surfaces roughness profiles and (b) six 3D-printed implants roughness profiles.*

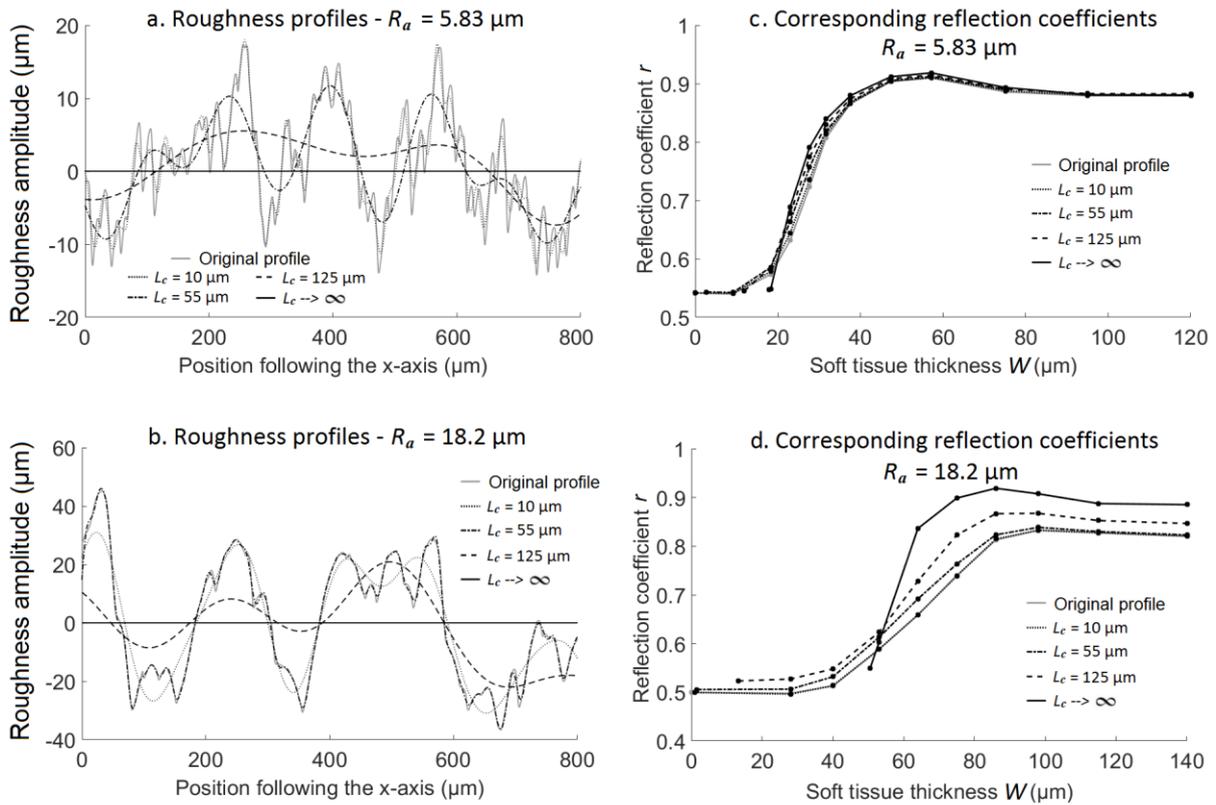

Fig. 5: *Roughness profiles of an implant surface with (a) $R_a = 5.83$ μm and (b) $R_a = 18.2$ μm together with the corresponding profiles filtered with different values of the cut-off*

lengths $L_c$. (c) and (d): Variation of the reflection coefficient r as a function of the soft tissue thickness W for the corresponding roughness profiles shown (a) and (b), respectively.

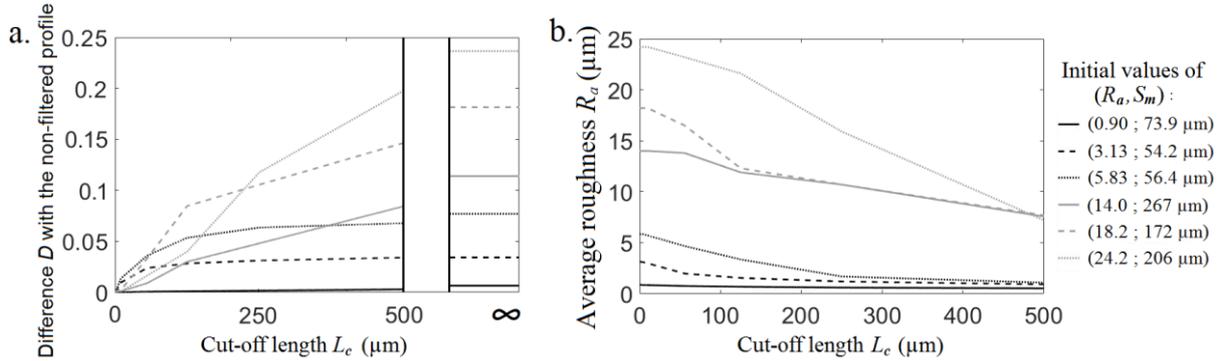

Fig. 6: Variation of (a) the difference D between the reflection coefficient of the filtered profiles and of the corresponding original profiles and (b) the average roughness $R_a$ of the filtered profiles as a function of the cut-off length $L_c$.

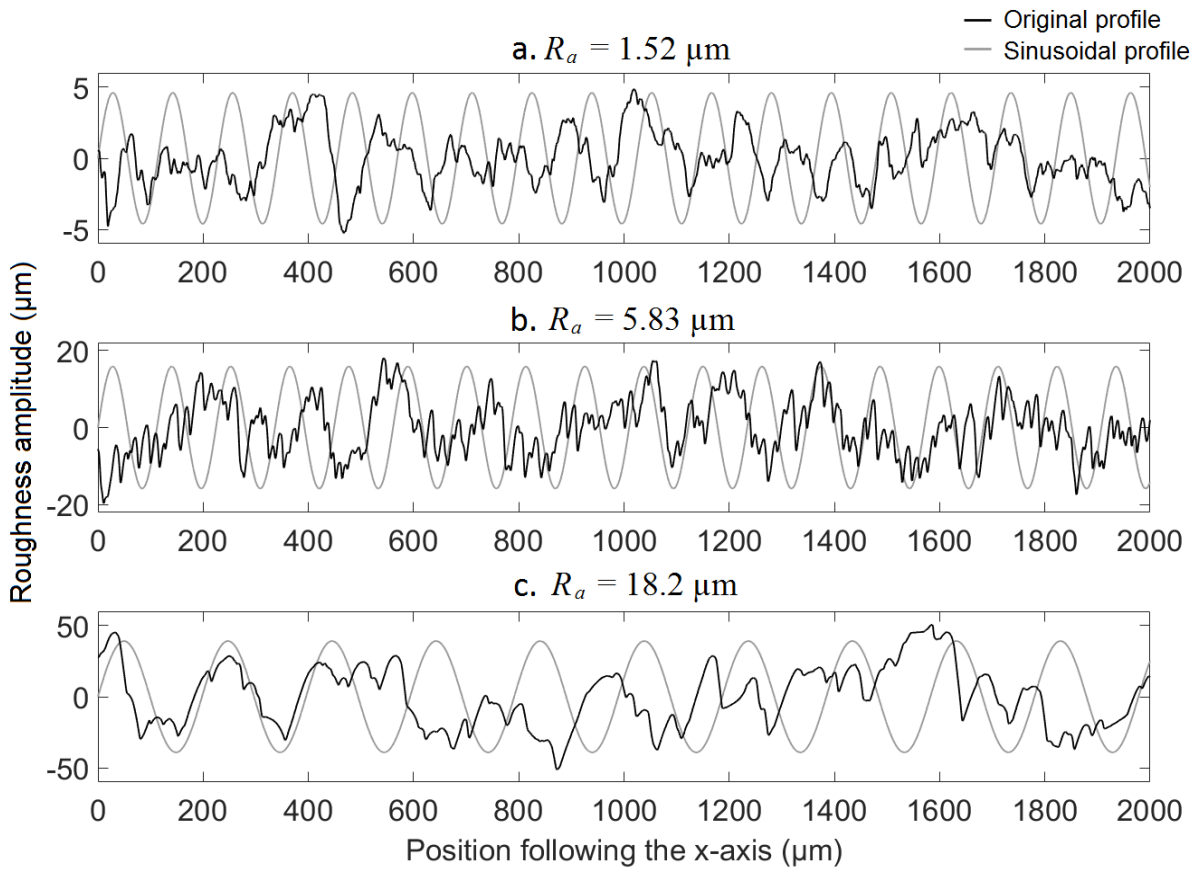

Fig. 7: Original roughness profiles of implants (black lines) with (a) $R_a = 1.52$ μm, (b) $R_a = 5.83$ μm, (c) $R_a = 18.2$ μm and corresponding optimized sinusoidal roughness profiles (grey lines).

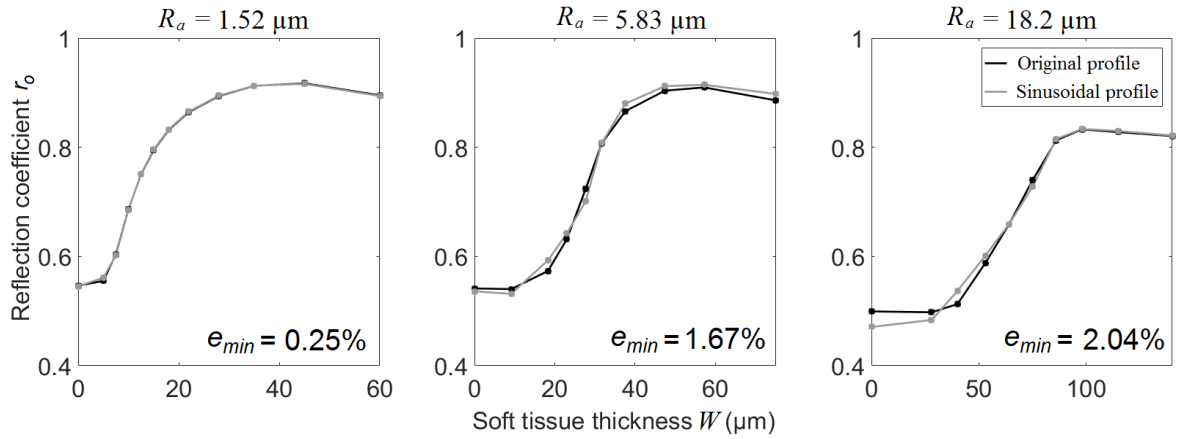

Fig. 8: *Variation of the reflection coefficient r as function of the soft tissue thickness W for roughness profiles of implants with (a) $R_a$ = 1.52 µm, (b) $R_a$ = 5.83 µm, (c) $R_a$ = 18.2 µm and for their corresponding optimized sinusoidal roughness profiles.*

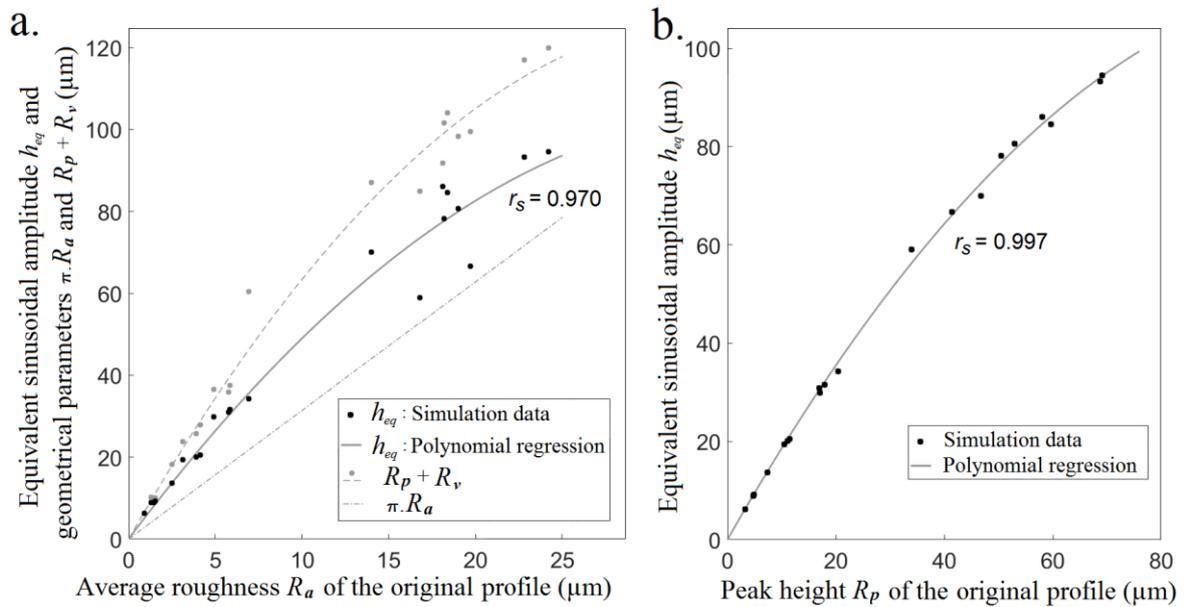

Fig. 9: *Variation of the optimized sinusoidal roughness amplitude $h_{eq}$ as a function of (a) the average roughness $R_a$ and (b) the maximum peak height $R_p$ of the implant. The solid lines and the equations correspond to the second order polynomial regression analysis of the variation of (a) $R_a$ and (b) $R_p$. The variation of $\pi.R_a$ and of $R_p + R_v$ as a function of $R_a$ are also represented in Fig. 9a.*

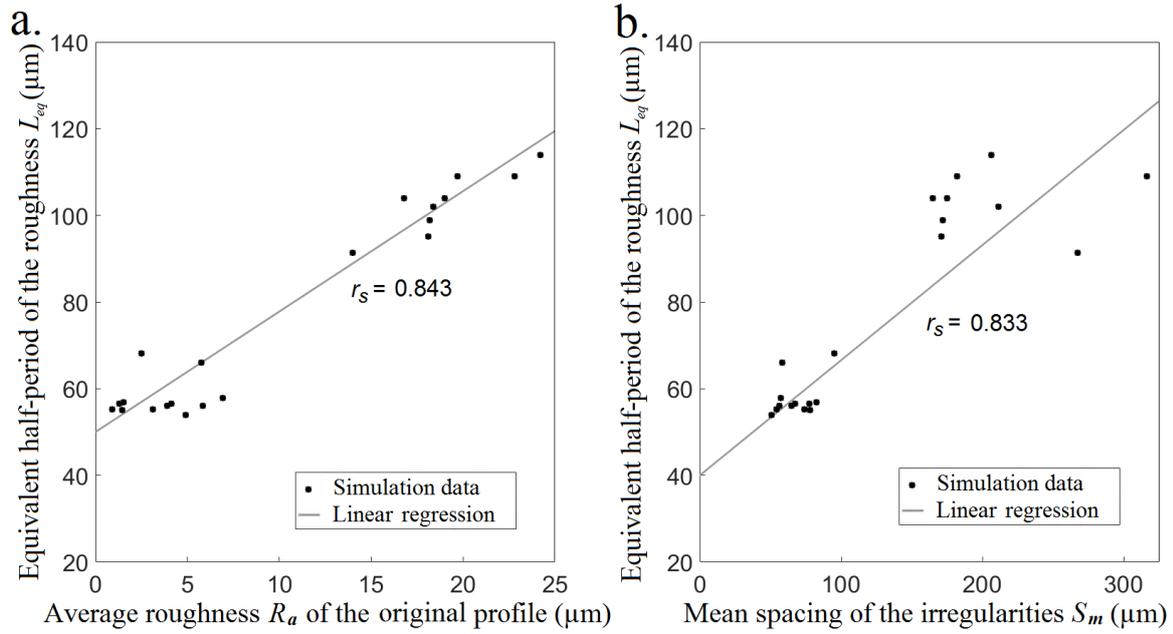

Fig. 10: *Variation of the optimized half-period of the roughness sinusoid $L_{eq}$ as a function of (a) the average roughness $R_a$ and (b) the mean spacing of irregularities $S_m$ of the implant. The solid lines and the equations correspond to a linear regression analysis.*

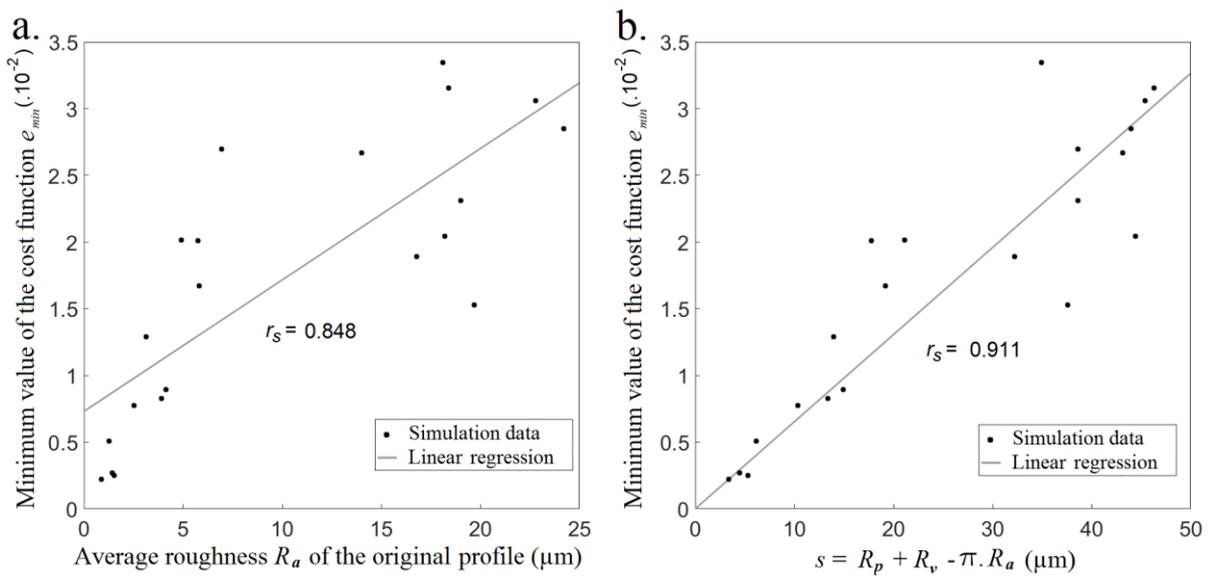

Fig.11: *Variation of the minimum value of the cost function $e_{min}$ as a function of (a) the average roughness of the implant $R_a$ and (b) $s = R_p + R_v - \pi.R_a$. The solid lines correspond to a linear regression analysis.*